\documentstyle[12pt,twoside,fleqn,espcrc1,epsfig]{article}
\def\bge{\begin{equation}}
\def\ene{\end{equation}}
\def\bg{\begin{eqnarray}}
\def\en{\end{eqnarray}}

\def\bge{\begin{equation}}
\def\ene{\end{equation}}
\def\bg{\begin{eqnarray}}
\def\en{\end{eqnarray}}

\def\ra{\rightarrow}

\def\What{\widehat{W}}
\def\Wzero{\widehat{W}_0}
\def\Wone{\widehat{W}_1}
\def\Wtwo{\widehat{W}_2}

\def\pslash{\not\!p}
\def\qslash{\not\!q}

\def\S0{{\Sigma^0}}

\def\X0{{\Xi^0}}

\newcommand{\AmS}{{\protect\the\textfont2
  A\kern-.1667em\lower.5ex\hbox{M}\kern-.125emS}}

\title{Deuteron Structure Functions in the Context of Few-Body Physics} 

\author{A. W. Thomas\address{Department of Physics and Mathematical
Physics, \\
and Special Research Centre for the Subatomic Structure of Matter, \\
University of Adelaide, Adelaide, Australia 5005}
and W. Melnitchouk\address{Department of Physics, University of
Maryland, \\ College Park, Maryland 20742, USA}}
\begin{document}
\maketitle
\vspace{-8cm}
\begin{flushright}
{\footnotesize 
Invited talk at the 15th Int. Conf. on Few-Body Problems in Physics \\
Groningen, The Netherlands, 22-26 July, 1997 \\
ADP-97-32/T265, DOE/ER/40762-129, UM PP 98-016}
\end{flushright}
\vspace{7cm}

\begin{abstract}
Few-body systems offer a unique challenge to those interested in
deep-inelastic scattering.
Using the deuteron as the simplest and most easily solvable nuclear system
we review the main physics issues in the valence and sea quark regimes.
For the former the key issue is the understanding of Fermi motion and
binding corrections, and our ability to extract the ratio $F_2^n/F_2^p$
as $x \rightarrow 1$.
The most recent analysis of deuteron data for $F_2^n/F_2^p$ is consistent
with perturbative QCD, but inconsistent with the assumptions common to
all standard parton distribution fits.
This should be taken into account in calculating event rates at HERA --- 
especially at large $x$ and $Q^2$.
The extraction of $g_1^n$ also presents an opportunity to refine the
description of nuclear effects in spin-dependent structure functions.
In the region of small $x$ one must correct for shadowing and meson 
exchange current effects to explore the flavor structure of sea, which is, 
in turn, deeply related to the chiral structure of the nucleon.
\end{abstract}

\section{INTRODUCTION}

The study of nuclear deep-inelastic scattering, particularly on the
deuteron, began in the 1970's, primarily as a source of information on
the neutron structure function.
There was an explosion of interest in the subject following the
announcement of the nuclear EMC effect in 1983 \cite{EMC}.
For an introduction to the ideas of deep-inelastic scattering and to
the nuclear EMC effect we refer to recent reviews \cite{ARN,GST}.
While the question of how the structure of the nucleon is modified in
medium \cite{GSRT} and how that may be investigated using deep-inelastic
scattering (DIS) \cite{SAI} is both timely and important, it is not
our prime concern here.
Instead, we concentrate on the deuteron as a source of information about
the structure of the neutron and as a laboratory for testing our
understanding of nuclear corrections such as binding, Fermi motion,
shadowing and meson exchange currents (MEC).
If we cannot understand these phenomena in a system as weakly bound as
the deuteron, then we shall certainly not understand them in any heavier
system.

In section 2 we briefly review the history of binding and Fermi motion
corrections, leading to a simple one-dimensional convolution of the
nucleon momentum distribution with the free nucleon structure function.
Recent work on the off-shell corrections to such a treatment is then
summarized, before we turn to a very interesting physics question,
namely the validity, or otherwise, of the perturbative QCD (PQCD) 
predictions for the asymptotic $d/u$ ratio in the proton.
This issue also seems to have consequences for the analysis of the
anomaly at large $x$ and $Q^2$ recently observed by ZEUS and H1 at HERA.
The extension to the spin dependent structure function of the neutron,
$g_1^n$, is also outlined.

In section 3 we turn to nuclear corrections at small $x$, the region
dominated by sea quarks.
The major surprise in this region in recent years was the discovery of
a significant violation of the Gottfried sum rule \cite{NMCO}, which
implied a major difference in the number of $\bar{u}$ and $\bar{d}$ 
quarks in the proton --- a conclusion supported in graphic fashion by
recent Drell-Yan data from Fermilab experiment E866 \cite{E866}.
The origin of this asymmetry seems to be the pion cloud of the nucleon 
\cite{THO83,SPTH}, which is a non-perturbative consequence of dynamical 
chiral symmetry breaking and which cannot be avoided {\em even} in the 
deep-inelastic regime.

\section{THE DEUTERON AT LARGE $x$}

The early work on DIS from the deuteron was aimed primarily at extracting
the neutron structure function \cite{KM,AW,BR}.
Amongst the many approaches to this problem we mention the light-front
treatment \cite{FS,KU} and the relativistic impulse approximation 
\cite{LG,BT}, involving the free nucleon structure function at a shifted 
value of $x$ or $Q^2$ \cite{DT,NW}.
A more phenomenological approach, developed by Frankfurt and Strikman
\cite{FS88}, attempts to derive the nuclear correction in the deuteron
by extrapolation from higher $A$ as a function of the ``effective density''
of the nucleus.
The experimental extraction of $F_2^n$ is usually made using either the
phenomenological effective density approach or the older, ``pre-EMC'' 
theoretical treatments.
We shall see that in particular the latter are not satisfactory,
and that taking into account the main lessons of the EMC effect
leads to a surprisingly large change in the conclusions concerning
the ratio $F_2^n/F_2^p$ in the region $x \rightarrow 1$.

The traditional impulse approximation treatment of the deuteron
structure function assumes a convolution of the free nucleon 
structure function, $F_2^N$, with the non-relativistic momentum
distribution $f_{N/D}$ of nucleons in the deuteron, calculated
in terms of a non-relativistic wave function of the deuteron and
its binding energy.
Although the binding energy is very small, the kinetic energy of the
recoiling nucleon plays a significant role in forcing the struck nucleon
further off-shell than one would usually expect --- as first pointed out
by Dunne and Thomas \cite{DT}:
\begin{equation}
f_{N/D}(y) \sim y \int d^3p \left| \psi_D(\vec{p}) \right|^2
	\delta\left(y - \frac{M_D - M - \vec{p}^2/2M}{M}\right),
\label{2}
\end{equation}
where $p$ is the off-shell nucleon momentum, $y$ is the light-cone
nucleon momentum fraction, and $\psi_D$ is the deuteron wave function.
Note that the ``flux-factor'', $y$, was omitted in early work, but as
noted by many authors \cite{FS88,MILL} it is important in practice.

\subsection{Covariant Formulation}
\label{covunpol}

In order to assess the theoretical reliability of the non-relativistic
impulse approximation one needs to go beyond the usual assumptions made
in the convolution approach \cite{CONV}.
In particular, the ingredients necessary for a covariant, relativistic
description are a covariant $DNN$ vertex with one of the nucleons
(the spectator to the hard collision) on-mass-shell, and an off-shell
photon--nucleon scattering amplitude (``off-shell nucleon structure
function'') $\widehat W$, the full structure of which was only recently
derived in Ref.\cite{MST1}.

The covariant $DNN$ vertex was written down many years ago by
Blankenbecler and Cook \cite{BC} and, at least in the rest frame
of the deuteron, can be expressed in terms of relativistic deuteron
wave functions \cite{BG,REL}.
The deuteron structure function can then be written in terms of the Dirac
trace of the product $(A_0 + \gamma^\mu A_{1 \mu}) \What$, where the
combination $(A_0 + \gamma^\mu A_{1 \mu})$ includes all of the
information on the deuteron vertex and (on-mass-shell) spectator nucleon,
summed over spin projections and can be calculated in terms of the
relativistic deuteron wave function.

The analysis of Ref.\cite{MST1} showed that the most general form of
the operator $\What$ (which is a $4\times 4$ matrix in Dirac space),
consistent with the discrete symmetries and gauge invariance, which 
contributes in the Bjorken limit is:
\begin{equation}
\What = \Wzero I + \Wone \pslash + \Wtwo \qslash,
\label{3}
\end{equation}
where the $\widehat W_i$ are functions of $p^2, q^2$ and $p \cdot q$
($q$ is the virtual photon four-momentum).
Thus, whereas in the free case the nucleon structure function involves
the combination:
\begin{equation}
{\rm Tr}[(\pslash + M) \What] \sim M \Wzero + M^2 \Wone + p \cdot q \Wtwo,
\label{4}
\end{equation}
the deuteron structure function involves:
\begin{equation}
{\rm Tr}[(A_0 + \gamma^\mu A_{1 \mu}) \What]
\sim A_0 \Wzero + p \cdot A_1 \Wone + q \cdot A_1 \Wtwo.
\label{5}
\end{equation}
Clearly then, even in the absence of Fermi motion one finds that
in general:
\begin{equation}
F_2^D \neq F_2^N,
\label{6}
\end{equation}
unless, as is usually implicitly assumed \cite{CONV}, only one of the
$\widehat W_i$ is non-zero \cite{MST1}.

Having established that, in principle, the structure function of the bound
nucleon cannot equal the structure function of the free nucleon, the
important question is how big the difference actually is in practice.
To estimate this, one can construct a simple model \cite{MST1} of the
so-called hand-bag diagram for an off-shell nucleon, in which the $N$-quark 
vertex is taken to be either a simple scalar or pseudo-vector, with the
parameters adjusted to reproduce the free nucleon structure functions
\cite{MSM}.
It is then possible to manipulate the exact deuteron structure function
into the form \cite{MST2}:
\begin{equation}
F_2^D = F_2^{D (\rm conv)} + \delta^{\rm (off)} F_2^D,
\label{7}
\end{equation}
where the convolution component:
\begin{equation}
F_2^{D (\rm conv)}(x,Q^2)
= \frac{1}{2} \sum_N \int^{M_D/M}_x dy\
\phi_{N/D}(y)\ F_2^N\left(\frac{x}{y},Q^2\right),
\label{8}
\end{equation}
now involves the relativistic momentum distribution of nucleons in the
deuteron \cite{MST2} (i.e. $\phi_{N/D}(y)$ depends on all components of
the relativistic deuteron wave function, $u^2 + w^2 + v_s^2 + v_t^2$
\cite{BG}).
Note that the argument of the free nucleon structure function
$F_2^N$ indicates that we need to find a quark with momentum $x$ in a 
nucleon with momentum $y$.
{}Furthermore, the ``non-convolution'' correction term
$\delta^{\rm (off)} F_2^D$ can be separated into two pieces associated 
with off-mass-shell corrections at the $DN$ and $\gamma^* N$ vertices 
\cite{MST2}.
\begin{figure}[hbt]
\begin{center}
\epsfig{file=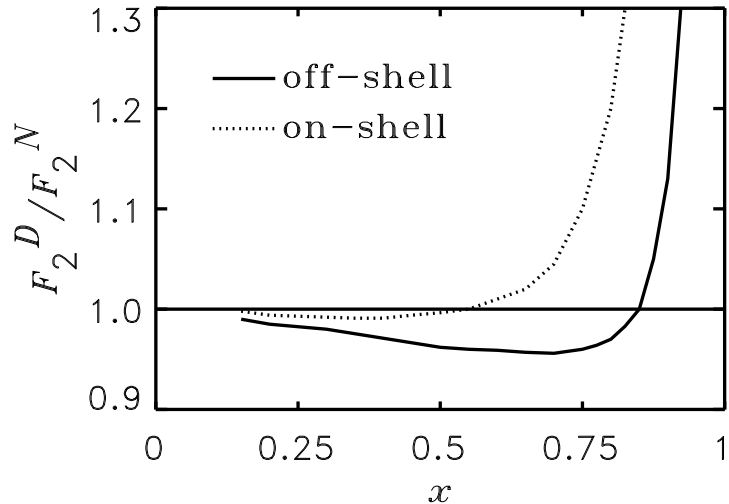,height=8cm}
\caption{$F_2^D/F_2^N$ ratio as a function of $x$ for the model of
	Refs.\protect\cite{MST1,MST2} (solid) which accounts for
	off-shell kinematics, and the on-shell model of
	Ref.\protect\cite{FS} (dotted) --- from Ref.\protect\cite{MST2}.}
\end{center}
\end{figure}

The result of the fully off-shell calculation from Ref.\cite{MST2}
is shown in Fig.1 (solid curve), where the ratio of the total
deuteron to nucleon structure functions ($F_2^D/F_2^N$) is plotted.
We also show the result of an on-mass-shell calculation from
Ref.\cite{FS} (dotted curve), which has been used in many previous 
analyses of the deuteron data \cite{EMC,WHIT}.
The most striking difference between the curves is the fact that the 
on-shell ratio has a very much smaller trough at $x \approx 0.3$, and 
rises faster above unity (at $x \approx 0.5$) than the off-shell curve, 
which has a deeper trough, at $x \approx 0.6-0.7$, and rises above unity 
somewhat later (at $x \approx 0.8$).

The behaviour of the full off-shell curve in Fig.1 is qualitatively
similar to that found by Uchiyama and Saito \cite{US}, Kaptari and Umnikov
\cite{KU}, and Braun and Tokarev \cite{BT}, who also used off-mass-shell
kinematics.
However, these authors did not include the (small) non-convolution
correction term, $\delta^{\rm (off)} F_2^D$.
The on-shell calculation \cite{FS}, on the other hand, was performed
in the infinite momentum frame where the nucleons are on-mass-shell.
The problem with this approach in the past has been the lack of
reliable deuteron wave functions in the infinite momentum frame,
and only recently have the first steps been taken in calculating
deuteron wave functions on the light-cone \cite{LIGHTFRONT}.
In practice one has usually made use of the non-relativistic,
$S$- and $D$-state deuteron wave functions calculated in the
deuteron rest frame, without accounting for the explicit binding
effects which should show up in the infinite momentum frame in
the form of additional Fock components (e.g., $NN$-meson(s) )
in the nuclear wave function.

Clearly, a smaller $D/N$ ratio at large $x$, as in Refs.\cite{MST1,MST2},
implies a larger neutron structure function in this region.
To estimate the size of the effect on the $n/p$ ratio requires one to
``deconvolute'' Eq.(\ref{8}) in order to extract $F_2^n$.
To study nuclear effects on the neutron structure function arising
from different models of the deuteron, one must eliminate any effects
that may arise from the extraction method itself.
Melnitchouk and Thomas \cite{MTNP} therefore use exactly the same 
extraction procedure as used in previous EMC \cite{EMC} and SLAC
\cite{WHIT} data analyses, namely the smearing (or deconvolution)
method discussed by Bodek {\em et al.} \cite{BODEK}.
{\em This method involves the direct use of the proton and deuteron 
data, without making any assumption concerning $F_2^n$ itself}.
For completeness we briefly outline the main ingredients in this method.

Firstly, one subtracts from the deuteron data, $F_2^D$, the additive, 
off-shell corrections, $\delta^{\rm (off)} F_2^D$, to give the convolution 
part, $F_2^{D (\rm conv)}$.
Then one smears the proton data, $F_2^p$, with the nucleon momentum 
distribution function $\phi_{N/D}(y)$ in Eq.(\ref{8}) to give 
$\widetilde{F}_2^p \equiv F_2^p/S_p$.
The smeared neutron structure function, $\widetilde{F}_2^n$, 
is then obtained from:
\begin{eqnarray}
\label{F2nsm}
\widetilde{F}_2^n &=& F_2^{D (\rm conv)} - \widetilde{F}_2^p.
\end{eqnarray}
Since the smeared neutron structure function is defined as
$\widetilde{F}_2^n \equiv F_2^n/S_n$, we can invert this to
obtain the structure function of a free neutron:
\begin{eqnarray}
\label{F2n}
F_2^n &=& S_n \left( F_2^{D (\rm conv)} - F_2^p/S_p \right).
\end{eqnarray}
In Eq.(\ref{F2n}), the proton smearing factor, $S_p$, can be computed
at each $x$ from the function $\phi_{N/D}(y)$, and a parameterization
of the $F_2^p$ data \cite{F2PAR}.
The neutron structure function may then be derived iteratively from
Eq.(\ref{F2n}).
Taking as a first guess $S_n = S_p$, the values of $F_2^n$ are smeared
by the function $\phi_{N/D}(y)$, and the results used to obtain a better
estimate for $S_n$.
The new value for $S_n$ is then used in Eq.(\ref{F2n}) to obtain
an improved estimate for $F_2^n$, and the procedure repeated until
convergence is achieved.
\begin{figure}[hbt]
\begin{center}
\epsfig{file=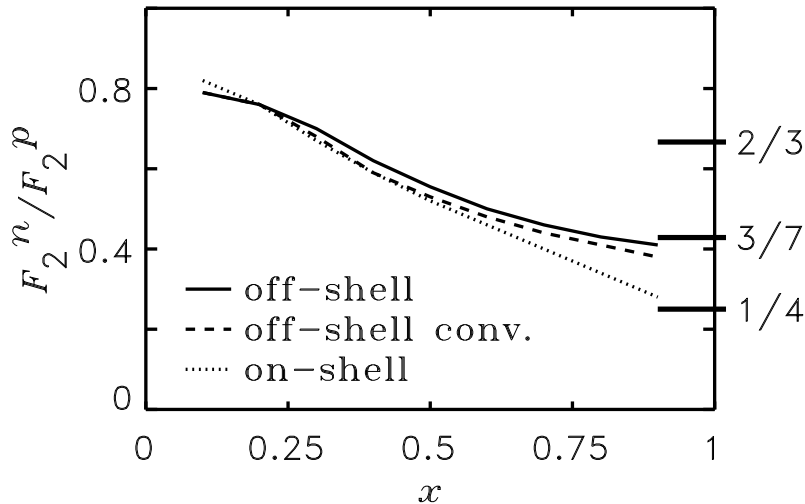,height=8cm}
\caption{The ratio $F_2^n/F_2^p$ as a function of $x$, for the off-shell 
	model (solid), off-shell model without the convolution-breaking term 
	(dashed), and the on-shell model (dotted).
	On the right-hand axis we have marked the $x \rightarrow 1$ limits of 
	the SU(6) symmetric model (2/3), and the predictions of the models of 
	Refs.\protect\cite{CLO73,CAR75} (1/4) and \protect\cite{FJ,BBS} (3/7).}
\end{center}
\end{figure}

The results of this procedure are presented in Fig.2, for both the
off-shell calculation \cite{MST2} (solid) and the on-shell model
\cite{FS} (dotted).
The increase in the off-shell ratio at large $x$ is a direct consequence
of the deeper trough in the $F_2^D/F_2^N$ ratio in Fig.1.
To illustrate the role of the non-convolution correction, 
$\delta^{\rm (off)} F_2^D$, we have also performed the analysis setting
this term to zero, and approximating $F_2^D$ by $F_2^{D\ {\rm (conv)}}(x)$.
The effect of this correction (dashed curve in Fig.2) appears minimal.
One can therefore attribute most of the difference between our analysis
and the earlier, ``pre-EMC'', on-shell results to kinematics --- since
both calculations involve essentially the same deuteron wave functions.

\subsection{A Test of Perturbative QCD}

The precise mechanism whereby the SU(6), spin-flavor symmetry of the parton 
distributions of the nucleon is broken is a question of fundamental 
importance in hadronic physics.
In a world of exact SU(6) symmetry, the wave function of a proton,
polarized say in the $+z$ direction, has the form \cite{CLO79}:
\begin{eqnarray}
\label{pwfn}
\left| p\uparrow \rangle \right.
&=& {1 \over \sqrt{2}} \left| u\uparrow (ud)_{S=0} \rangle \right. \
 +\ {1 \over \sqrt{18}} \left| u\uparrow (ud)_{S=1} \rangle \right. \
 -\ {1 \over 3}  \left| u\downarrow (ud)_{S=1} \rangle \right. \ \nonumber \\
& &
 -\ {1 \over 3}  \left| d\uparrow (uu)_{S=1} \rangle \right. \
 -\ {\sqrt{2} \over 3} \left| d\downarrow (uu)_{S=1} \rangle \right. ,
\end{eqnarray}
where the subscript $S$ denotes the total spin of the two-quark component.
In this limit, the nucleon and $\Delta$ isobar would be degenerate.
In deep-inelastic scattering, exact SU(6) symmetry would be manifested in 
equivalent shapes for the valence quark distributions of the proton, which 
would be related simply by $u_V(x) = 2 d_V(x)$ for all $x$.
For the neutron to proton structure function ratio this would imply:
\begin{eqnarray}
{ F_2^n \over F_2^p }
&=& {2 \over 3}\ \ \ \ \ \  ; \ \ \ \ {\rm SU(6)\ symmetry}.
\end{eqnarray}

Of course, Nature does not respect exact SU(6) spin-flavor symmetry.
The nucleon and $\Delta$ masses are split by some 300 MeV, and empirically
the $d$ quark distribution is softer than the $u$.
The correlation between the mass splitting in the {\bf 56} baryons and the 
large-$x$ behavior of $F_2^n/F_2^p$ was observed some time ago by Close 
\cite{CLO73} and Carlitz \cite{CAR75}.
Based on phenomenological \cite{CLO73} and Regge \cite{CAR75} arguments,
the breaking of the symmetry in Eq.(\ref{pwfn}) was argued to arise from
a suppression of the ``diquark'' configurations having $S=1$ relative to
the $S=0$ configuration. 
Such a suppression is, in fact, quite natural if one observes that whatever
mechanism leads to the observed $N-\Delta$ splitting (e.g. color-magnetic 
force, instanton-induced interaction, pion exchange), necessarily acts to 
produce a mass splitting between the possible spin states of the spectator   
pair, $(qq)_S$, with the $S=1$ state heavier than the $S=0$ state by some
200 MeV \cite{CT}.
{}From Eq.(\ref{pwfn}), a dominant scalar valence diquark component of the 
proton suggests that in the $x \rightarrow 1$ limit $F_2^p$ is essentially 
given by a single quark distribution (i.e. the $u$), in which case:
\begin{eqnarray}
{ F_2^n \over F_2^p }
&\rightarrow& { 1 \over 4 }, \ \ \ \ \
{ d \over u } \rightarrow 0\ \ \ \ \
; \ \ \ \ S=0\ {\rm dominance}.
\end{eqnarray}
This expectation has, in fact, been built into all phenomenological fits
to the parton distribution data.

An alternative suggestion, based on perturbative QCD, was originally
formulated by Farrar and Jackson \cite{FJ}.
There it was argued that the exchange of longitudinal gluons, which are
the only type permitted when the spin projections of the two quarks in
$(qq)_S$ are aligned, would introduce a factor $(1-x)^{1/2}$ into the
Compton amplitude --- in comparison with the exchange of a transverse
gluon between quarks with spins anti-aligned.
In this approach the relevant component of the proton valence wave function
at large $x$ is that associated with states in which the total ``diquark''
spin {\em projection}, $S_z$, is zero.
Consequently, scattering from a quark polarized in the opposite direction
to the proton polarization is suppressed by a factor $(1-x)$ relative to
the helicity-aligned configuration.

A similar result is also obtained in the treatment of Brodsky {\em et al.} 
\cite{BBS} (based on counting-rules), where the large-$x$ behavior of the 
parton distribution for a quark polarized parallel ($\Delta S_z = 1$) or 
antiparallel ($\Delta S_z = 0$) to the proton helicity is given by:
$q^{\uparrow\downarrow}(x) = (1~-~x)^{2n - 1 + \Delta S_z}$,
where $n$ is the minimum number of non-interacting quarks (equal to 2 for 
the valence quark distributions).
In the $x \rightarrow 1$ limit these arguments, based on PQCD, suggest:
\begin{eqnarray}
{ F_2^n \over F_2^p }
&\rightarrow& {3 \over 7}, \ \ \ \ \
{ d \over u } \rightarrow { 1 \over 5 }\ \ \ \ \
; \ \ \ \ S_z=0\ {\rm dominance}.
\end{eqnarray}
Note that the $d/u$ ratio {\em does not vanish} in this case.
Clearly, if one is to understand the dynamics of the nucleon's quark 
distributions at large $x$, it is imperative that the consequences of
these models be tested experimentally.

As explained above, our analysis of the SLAC data points \cite{WHIT,GOMEZ} 
involves no assumption whatsoever about $F_2^n$, the only input is the
nucleon momentum density in the deuteron, $\phi_{N/D}(y)$.
The results of this reanalysis are shown in Fig.3, at an average value of
$Q^2 \approx 12$ GeV$^2$.
The data represented by the open circles have been extracted with the
on-shell deuteron model of Ref.\cite{FS}, while the filled circles were
obtained using the off-shell model of Refs.\cite{MST1,MST2}.
Most importantly, the $F_2^n/F_2^p$ points obtained with the off-shell
method appear to approach a value broadly consistent with the
Farrar-Jackson \cite{FJ} prediction of 3/7, whereas the data previously
analyzed in terms of the on-shell formalism produced a ratio tending to
the lower value of 1/4.
\begin{figure}[hbt]
\begin{center}
\epsfig{file=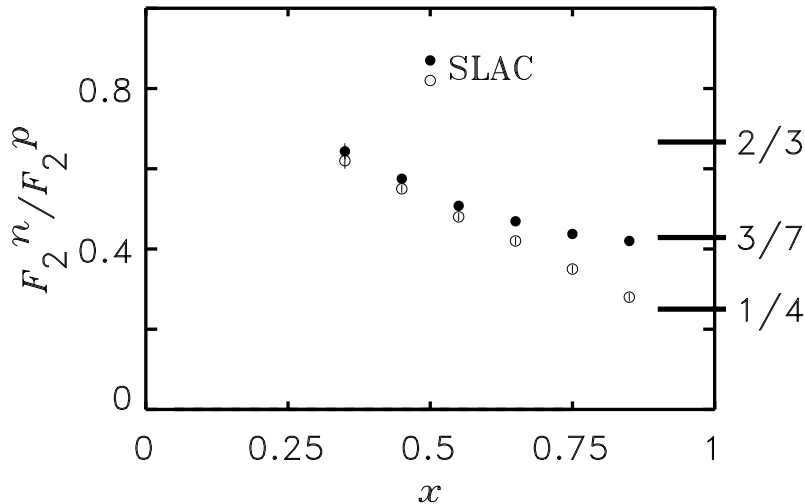,height=8cm}
\caption{Deconvoluted $F_2^n/F_2^p$ ratio extracted from the
	SLAC $p$ and $D$ data \protect\cite{WHIT,GOMEZ}	using
	the model of Ref.\protect\cite{MST1,MST2} (solid circles)
	and Ref.\protect\cite{FS} (open circles).}
\end{center}
\end{figure}

The $d/u$ ratio, shown in Fig.4, is obtained by inverting $F_2^n/F_2^p$
in the valence quark dominated region.
The points extracted using the off-shell formalism (solid circles) are
again significantly above those obtained previously with the aid of the
on-shell prescription.
In particular, they indicate that the $d/u$ ratio may actually approach
a {\em finite} value in the $x \rightarrow 1$ limit, contrary to the
expectation of the model of Refs.\cite{CLO73,CAR75}, in which $d/u$
tends to zero.
Although it is {\em a priori} not clear at which scale these model
predictions should be valid, for the values of $Q^2$ corresponding
to the analyzed data the effects of $Q^2$ evolution are minimal.
\begin{figure}[hbt]
\begin{center}
\epsfig{file=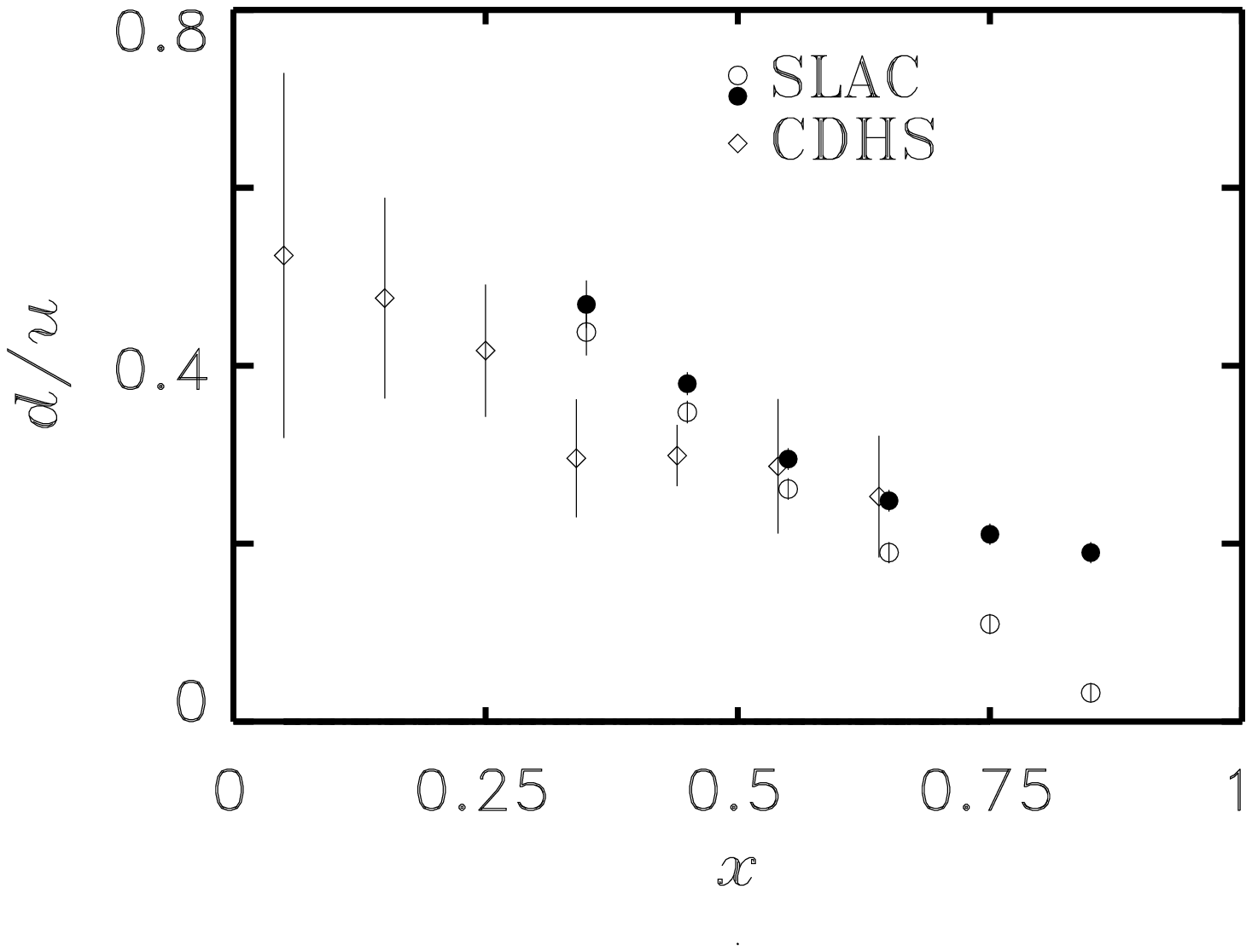,height=8cm}
\caption{Extracted $d/u$ ratio (see Fig.2).
	Also shown for comparison is the ratio extracted from neutrino
	measurements by the CDHS collaboration \protect\cite{CDHS}.}
\end{center}
\end{figure}

Naturally, it would be preferable to extract $F_2^n$ at large $x$
without having to deal with uncertainties in the nuclear effects.
In principle this could be achieved by using neutrino and
antineutrino beams to measure the $u$ and $d$ distributions in
the proton separately, and reconstructing $F_2^n$ from these.
Unfortunately, as seen in Fig.4, the neutrino data from the
CDHS collaboration \cite{CDHS} do not extend out to very large
$x$ ($x > 0.6$), and at present cannot discriminate between
the different methods of analyzing the electron--deuteron data.

\subsection{Relevance of the $d/u$ Ratio to the HERA Anomaly}

The H1 and ZEUS experiments at HERA have recently produced a small
number of events at enormously high $Q^2$ which have generated
tremendous theoretical interest \cite{H1,ZEUS}.
For $Q^2 > 10,000 $GeV$^2$ and $x > 0.45$ the valence parton distributions
are calculated to drop dramatically.
The HERA anomaly is essentially the excess of observed ``neutral current''
events (i.e., events of the type $e^+ p \ra e^+ X$) 
over expectations by roughly
an order of magnitude.
Many exotic explanations of this excess have already been suggested, indeed,
the number of possibilities currently exceeds the number of events.
However, before the new physics can be worked out one must be sure that the 
input parton distributions used to estimate ``background'' rates are reliable.

One glaring problem with the current treatment of the partonic 
``background'' is that {\em all} of the standard distributions used are
constructed to satisfy $d/u \ra 0$ as $x \ra 1$ at low-$Q^2$.
As we have seen, the recent re-analysis of the deuteron data leads to a
$d/u$ ratio which appears to be consistent with the prediction of PQCD
that $d/u \ra 1/5$ as $x \ra 1$.
In the light of this result it is not only vital to find alternative,
more direct measurements of $d/u$ at large $x$, but those generating
standard sets of parton distributions should at the very least present
alternative parameter sets consistent with the new analysis of the
deuteron data.
Until parameter sets are constructed which are consistent with
$d/u \ra 1/5$ as $x \ra 1$, at $Q^2 \sim 10$ GeV$^2$, one cannot be
sure of the reliability of ``background'' rate estimates at the extreme
values of $Q^2$ and $x$ being probed at HERA.
Certainly one cannot hope to probe other exciting aspects of hadron
structure, such as intrinsic charm and beauty \cite{BROD,KLT,MT_HERA},
until the $d/u$ issue has been resolved.

\subsection{Spin Dependent Structure Functions}

The spin structure functions of the nucleon, $g_1^{p(n)}$, are of
tremendous interest at present.
Experimentally, $g_1$ is proportional to the difference of DIS cross
sections for $ep$ scattering with beam and target helicities aligned
and anti-aligned \cite{CHENG}. 
Within the parton model it may be written in terms of the parton helicity 
(loosely ``spin'') distributions,
$\Delta q(x) = [q^{\uparrow} - q^{\downarrow}
	       + \bar{q}^{\uparrow} - \bar{q}^{\downarrow}]$,
with $q^{\uparrow (\downarrow)}$ the number density of quarks with
helicity parallel (anti-parallel) to the helicity of the target proton:
\bge
g_1^p(x) = \frac{1}{2} \sum_q e^2_q \Delta q(x).
\label{eq:3.1}
\ene

Intense interest in the spin structure functions began in 1988 when EMC
announced a large violation of the Ellis-Jaffe sum rule \cite{EMC_SPIN}, 
which relates $\Gamma_p(Q^2) \equiv \int^1_0 g_1^p(x,Q^2) dx$ to the
isovector and octet axial-vector coupling constants, $g_A^{(3)}$ and
$g_A^{(8)}$.
The failure of this sum rule, which is not a rigorous consequence of QCD,
led to questions about the Bjorken sum rule, which relates
$\Gamma_p - \Gamma_n$ to $g_A^{(3)}/6$ (modulo QCD radiative corrections
\cite{VER}) and {\em is} a strict consequence of QCD.
To determine $\Gamma_n$ one must measure $g_1^n(x)$, which requires a
polarized nuclear target such as $^3$He or D.
At present, all neutron data extracted from the deuteron are obtained
by applying a simple, non-relativistic prescription to correct $g_1^D$ 
for the $D$-state component (probability $\omega_D$) of the deuteron
wave function \cite{DPOL}:
\bge
g_1^n(x)
= \left(1 - \frac{3}{2} \omega_D\right)^{-1} g_1^D(x) - g_1^p(x).
\label{eq:3.3}
\ene
As we explain below, exactly the same techniques as described in the
previous section may be used to test the accuracy of Eq.(\ref{eq:3.3}).

While most interest has been focussed on the issue of sum rules,
we stress that the shapes of $g_1^p(x)$ and $g_1^n(x)$ contain even
more important information.
For example, the same arguments that led to different conclusions
about the behaviour of $d/u$ as $x \ra 1$ also give quite different 
predictions for $g_1^p$ and $g_1^n$ as $x \rightarrow 1$, namely 1/4
and 3/7, respectively.
Quite interestingly, while the ratio of the 
polarized to unpolarized $u$
quark distributions is predicted to be the same in the two models:
\begin{eqnarray}
\label{Deltau}
{ \Delta u \over u }
&\rightarrow& 1\ \ \ \ ; \ \ \ \ S=0\ or\ S_z=0\ {\rm dominance},
\end{eqnarray}
the results for the $d$-quark distribution ratio differ even in sign:
\begin{eqnarray}
{ \Delta d \over d }
&\rightarrow& - {1 \over 3}\ \ \ \ ; \ \ \ \ S=0\ {\rm dominance},\\
&\rightarrow& 1\ \ \ \ ; \ \ \ \ S_z=0\ {\rm dominance}.
\label{Deltad}
\end{eqnarray}

Using the same techniques described in section \ref{covunpol} for the
unpolarized case, Melnitchouk, Piller and Thomas \cite{MPT} derived the
most general, antisymmetric, Dirac tensor operator of twist 2 for an
off-mass-shell nucleon (see also \cite{KMPW}):
\bge
\widehat{G}^{\mu\nu} = i \epsilon^{\mu\nu\alpha\beta} q_{\alpha}
\left[ p_\beta ( \pslash \gamma_5 \widehat G_p
	       + \qslash \gamma_5 \widehat G_q )
     + \gamma_\beta \gamma_5 \widehat G_\gamma 
\right].
\label{eq:G}
\ene
Once again, one finds that three functions, $\widehat G_i$, can be
constructed in terms of scalar and pseudo-scalar vertices.
However, in this case there is a new feature, {\em the function
$\widehat G_p$ does not contribute for a free nucleon}, whereas it
does contribute in a nucleus.
\begin{figure}[hbt]
\begin{center}
\epsfig{file=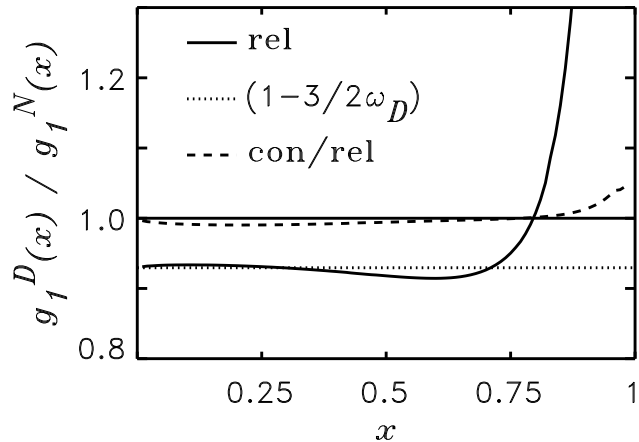,height=7cm}
\caption{Ratio of deuteron and nucleon structure functions in the
	full model (solid), and with a constant depolarization factor
	corresponding to $\omega_D = 4.7\%$ (dotted line).
	The dashed curve is the ratio of $g_1^D$ calculated via
	convolution to $g_1^D$ calculated in the relativistic model
	--- from Ref.\protect\cite{MPT}.}
\end{center}
\end{figure}

The spin-dependent deuteron structure function is given by the trace of
$\widehat{G}^{\mu\nu}$ with a spin-dependent $ND$ amplitude \cite{MPT},
which can be evaluated using the relativistic $DNN$ vertex of Ref.\cite{BG}.
Surprisingly, Fig.5 shows that the ratio of the convolution approximation
to the fully off-shell calculation (the dashed curve) is even closer to
unity in this case than in the spin-independent case.

The comparison between the solid and dotted curves in Fig.5 shows that
Eq.(\ref{eq:3.3}) is reliable at the 2\% level for $x$ below 0.7.
However, the excellent agreement in this region between (\ref{eq:3.3})
and the full calculation relies on a knowledge of the deuteron $D$-state
probability.
As shown in Ref.\cite{MPT}, a change of $\omega_D$ by 2\% (e.g. from 4\%
to 6\%) leads to an error of order 10\% or more in $g_1^n$.
For $x > 0.7$, on the other hand, the approximation (\ref{eq:3.3})
fails dramatically.
This will be extremely important when testing the predictions of PQCD
for the $x \ra 1$ behavior of the polarized distributions in
Eqs.(\ref{Deltau})--(\ref{Deltad}).

The use of a polarized $^3$He target reduces the errors arising from
the extraction, but in that case we have, as yet, no analogue of the
relativistic vertex which permitted an estimate of off-shell corrections
in the deuteron case.
As the binding energy and density of the target nucleus increases we
expect these corrections to increase as well.
In particular, while the new structure function $\widehat G_p$ contributes
only at order $(v/c)^3$ and is negligible for the deuteron, one cannot yet
discount the possibility of $\widehat G_p$ playing a significant role in
nuclei heavier than the deuteron.

\section{THE DEUTERON AT SMALL $x$}

At decreasing values of Bjorken-$x$, as sea quarks begin to dominate,
the impulse approximation should eventually break down.
One expects that coherent, multiple scattering effects become important 
when the characteristic time scale, $1/Mx$, of the DIS process becomes 
larger than the typical average distance between nucleon in the nucleus.
This occurs typically in the region $x < 0.1$.
Such effects are certainly seen in the low-$x$ depletion of the nuclear
EMC ratio in heavy nuclei.
Any shadowing in the deuteron should therefore produce a depletion in the 
ratio $F_2^D/F_2^N$ at small $x$ \cite{KWBD,KW,BK,MTD,MTA,OTHERD}.

A deviation from unity of $F_2^D/F_2^N$ at low $x$ is highly relevant
for understanding the violation of the Gottfried sum rule, which measures 
the integrated difference between the proton and neutron structure
functions.
The violation, discovered by the New Muon Collaboration in 1990 \cite{NMC},
was interpreted as a sizable difference between $\bar{d}$ and $\bar{u}$
in the proton, but relies on an accurate determination of nuclear
corrections to $F_2^D$.

An excess of $\bar{d}$ over $\bar{u}$ cannot be understood in terms of PQCD 
--- it is a property of the ``intrinsic sea'' of the nucleon.
We briefly review the interpretation of this excess, including the widely
accepted explanation in terms of the pion cloud of the nucleon, first
suggested in 1983 \cite{THO83,CBM}. 
As the pion cloud of the nucleon is an unavoidable consequence of dynamical 
symmetry breaking in QCD this experimental discovery offers important insight 
into the realization of non-perturbative QCD in the structure of the nucleon.

\subsection{Nuclear Shadowing}

The rescattering of a virtual photon from both nucleons in the deuteron
is usually described within the Glauber multiple scattering formalism.
Corrections associated with a relativistic deuteron wave function amount 
to a few percent out of a shadowing correction to $F_2^D$ that is only a 
few percent in total, and hence can be neglected.
At small $x$, nuclear binding and Fermi motion corrections can be neglected,
and the total deuteron structure function written as:
\begin{eqnarray}
F_2^D(x,Q^2)
&\approx& {1 \over 2}
\left( F_2^p(x,Q^2) + F_2^n(x,Q^2) + \delta^{\rm (shad)} F_2^D(x,Q^2)
\right).
\end{eqnarray}
In modeling the shadowing correction, $\delta^{\rm (shad)} F_2^D$,
our approach is to take a two-phase model, similar to that of 
Kwiecinski and Badelek \cite{KWBD,KW,BK}.
At high virtuality the interaction of the virtual photon with the deuteron
is parameterized in terms of diffractive scattering through the double and
triple Pomeron, as well as scattering from exchanged mesons in the deuteron.
On the other hand, at low virtuality it is most natural to apply
a vector meson dominance (VMD) model, in which the virtual photon
interacts with the nucleons via its hadronic structure, namely the 
$\rho^0$, $\omega$ and $\phi$ mesons.
The latter contribution vanishes at sufficiently high $Q^2$, but  
for $Q^2 < 1$ GeV$^2$ it is, in fact, responsible for the majority 
of the dependence on $Q^2$.

For the diffractive component, Pomeron ($I\!\!P$) exchange between the 
projectile and tboth constituent nucleons models the interaction of 
partons from different nucleons within the deuteron.
Assuming factorization of the diffractive cross section, the shadowing 
correction (per nucleon) to the deuteron structure function from 
$I\!\!P$-exchange is written as a convolution of the Pomeron structure 
function, $F_2^{I\!P}$, with a distribution function $f_{I\!P}$, 
describing the number density of exchanged Pomerons:
\begin{eqnarray}
\label{dFAP}
\delta^{(I\!P)} F_2^D(x,Q^2)
&=& \int_{y_{min}}^2\ dy\ 
f_{I\!P}(y)\ F_2^{I\!P}(x_{I\!P},Q^2),
\end{eqnarray}
where, treating the deuteron non-relativistically, 
\cite{KWBD,KW,BK,MTD,MTA}:
\begin{eqnarray}
f_{I\!P}(y)
&=& - \frac{\sigma_{pp}}{8 \pi^2}\ \frac{1}{y}
\int d^2{\bf k}_T\ S_{D}({\bf k}^2).
\end{eqnarray}
Here, $y = x (1+M_X^2/Q^2)$ is the light-cone momentum fraction carried 
by the Pomeron ($M_X$ is the mass of the diffractive hadronic debris), 
and $x_{I\!P} = x/y$ is the momentum fraction of the Pomeron carried by 
the struck quark in the Pomeron.
The deuteron form factor, $S_{D}({\bf k}^2)$, is given in terms of the 
coordinate space wave functions:
\begin{eqnarray}
S_{D}({\bf k}^2)
&=& \int_{0}^{\infty} dr \left| \psi_D(r) \right|^2
j_{0}(|{\bf k}| r),
\end{eqnarray}
where $j_{0}$ is a spherical Bessel function.
Within experimental errors, the factorization hypothesis, as well as the 
$y$ dependence of $f_{I\!P}(y)$ \cite{KWBD,KW,BK,MTD,MTA}, is consistent
with the recent HERA data \cite{HERA} obtained from observations of 
large rapidity gap events in diffractive $ep$ scattering.
These data also confirm previous findings that the Pomeron structure function
contains both a hard and a soft component:\
$ F_2^{I\!P}(x_{I\!P},Q^2)
= F_2^{I\!P ({\rm hard})}(x_{I\!P},Q^2)
+ F_2^{I\!P ({\rm soft})}(x_{I\!P},Q^2)$.
The hard component of $F_2^{I\!P}$ is generated from an explicit $q\bar q$
component of the Pomeron, and has an $x_{I\!P}$ dependence given by\
$x_{I\!P} (1-x_{I\!P})$ \cite{DOLA}, in agreement with the recent 
diffractive data \cite{HERA}.
The soft part, which is driven at small $x_{I\!P}$ by the triple-Pomeron 
interaction \cite{KWBD}, has an $x_{I\!P}$ dependence typical of a 
sea-quark distribution, with normalization fixed by the triple-Pomeron 
coupling constant.

The dependence of $F_2^{I\!P}$ on $Q^2$ \cite{KW}, at large $Q^2$, leads 
to a weak (logarithmic) $Q^2$ dependence for the shadowing correction 
$\delta^{(I\!P)} F_2^D$.
The low-$Q^2$ extrapolation of the $q\bar q$ component is parameterized by 
applying a factor $Q^2/(Q^2 + Q_0^2)$, where $Q_0^2 \approx 0.485$ GeV$^2$
may be interpreted as the inverse size of partons inside the virtual 
photon \cite{DLQ}.
For the nucleon sea quark densities relevant for $F_2^{I\!P ({\rm soft})}$
we use the recent parametrization of Donnachie and Landshoff, which 
includes a low-$Q^2$ limit consistent with the real photon data, in 
which case the total Pomeron contribution
$\delta^{(I\!P)} F_2^D \rightarrow 0$ as $Q^2 \rightarrow 0$ \cite{DLQ}.

An adequate description of shadowing at small $Q^2$ requires 
a higher-twist mechanism, such as vector meson dominance.
VMD is empirically based on the observation that some aspects
of the interaction of photons with hadronic systems resemble
purely hadronic interactions.
In terms of QCD this is understood in terms of a coupling of the
photon to a correlated $q\bar q$ pair of low invariant mass,
which may be approximated as a virtual vector meson.
One can then estimate the amount of shadowing in terms of the
multiple scattering of the vector meson using Glauber theory.
The corresponding correction (per nucleon) to the nuclear
structure function is:
\begin{eqnarray}
\label{dFAV}
\delta^{(V)} F_2^D(x,Q^2)
&=& \frac{ Q^{2} }{ \pi }
\sum_V
{ \delta\sigma_{VD} \over f_V^2 (1 + Q^2/M_V^2)^2 },
\end{eqnarray}
where 
\begin{eqnarray}
\delta \sigma_{VD}
&=& -\frac{ \sigma_{VN}^{2} }{ 8 \pi^{2} }
\int d^2{\bf k}_T S_{D}({\bf k}^2)    
\end{eqnarray}
is the shadowing correction to the vector meson--nucleus cross section, 
$f_V$ is the photon--vector meson coupling strength, and $M_V$ is the 
vector meson mass ($V = \rho^0, \omega, \phi$) are important at low $Q^2$.
The vector meson propagators in Eq.(\ref{dFAV}) lead to a strong $Q^2$ 
dependence of $\delta^{(V)} F_2^D$, which peaks at $Q^2 \sim 1$ GeV$^2$.
{}For $Q^2 \rightarrow 0$ and fixed $x$, $\delta^{(V)} F_2^D$ disappears 
due to the vanishing of the total $F_2^D$.
{}Furthermore, since this is a higher twist effect, shadowing in the VMD
model dies off quite rapidly between $Q^2 \sim 1$ and 10 GeV$^2$, so
that for $Q^2 > 10$ GeV$^2$ it is almost negligible --- leaving only the
diffractive term, $\delta^{(I\!P)} F_2^D$.
(Note that at fixed $\nu$, for decreasing $Q^2$ the ratio $F_2^D/F_2^p$
approaches the photoproduction limit.)

\subsection{Meson Exchange Currents}

Meson exchange currents (MEC) are familiar in few-body physics.
In the context of DIS they have been recognized as a potentially important,
many-body correction, which does indeed scale, since the possible 
enhancement of the nuclear pion field was proposed \cite{CLLS,ET}
as an explanation of the nuclear EMC effect.
In the deuteron case, Kaptari {\em et al.} have suggested \cite{KAP} that
MEC may lead to some {\em antishadowing} corrections to $F_2^D(x)$.
The total contribution to the deuteron structure function from meson 
exchange is:
\begin{eqnarray}
\delta^{(M)} F_2^D(x,Q^2)
&=& \sum_M
\int_x dy\ f_M(y)\ F_2^M(x/y,Q^2), 
\end{eqnarray}
where $M = \pi, \rho, \omega, \sigma$.
The virtual meson structure function, $F_2^M$, one approximates by the 
(real) pion structure function, for which data has been obtained through 
the Drell-Yan process.
The exchange-meson distribution functions, $f_M(y)$, are obtained from 
the non-relativistic reduction of the nucleon--meson interaction given 
in Refs.\cite{MTD,KAP}.
In practice pion exchange is the dominant process, and gives a positive 
contribution to $\delta^{(M)} F_2^D$.
The exchange of the fictitious $\sigma$ meson (which represents  
correlated $2 \pi$ exchange) also gives rise to antishadowing
for small $x$.
Vector meson exchange cancels some of this antishadowing, however the
magnitude of this is smaller.
In fact, for not too hard meson--nucleon vertices all contributions
other than that of the pion are essentially negligible.

\subsection{Neutron Structure Function at Small $x$}

In the region $x < 0.1$ the magnitude of the (negative) Pomeron/VMD  
shadowing is larger than the (positive) meson-exchange contribution,
so that the total
$ \delta^{\rm (shad)} F_2^D
= \delta^{(I\!P)} F_2^D
+ \delta^{(V)} F_2^D
+ \delta^{(M)} F_2^D $
is negative.
On the other hand, at slightly larger $x$ ($\sim$ 0.1--0.2), there is 
a very small amount of antishadowing, which is due mainly to the VMD 
and pion-exchange contributions.
In the kinematic region covered by NMC, $x > 0.004$, $Q^2=4$ GeV$^2$
\cite{N_D}, the overall effect on the shape of the neutron structure 
function is a 1--2\% {\em increase} in $F_2^n/F_2^{n \rm (bound)}$ for
$x < 0.01$, where $F_2^{n (\rm bound)} \equiv F_2^D - F_2^p$.

The presence of shadowing in the deuteron would be confirmed through
observation of a deviation from unity in the $F_2^D/F_2^p$ structure
function ratio in the kinematic region where Regge theory is expected
to be valid.
Although the exact value of $x$ below which the proton and (free) neutron
structure functions become equivalent is not known, it is expected that 
at low enough $x$, $F_2^p \rightarrow F_2^n$, in which case
$F_2^D/F_2^p \rightarrow 1 + \delta^{\rm (shad)} F_2^D/2F_2^p$.
Subtracting the shadowing correction from the deuteron data should
then yield $F_2^p - F_2^n \ra 0$ as $x \ra 0$.
While for the lowest NMC data point it may be debatable whether the
Regge region is reached, the E665 Collaboration \cite{E_D} has taken
data to very low $x$, $x \sim 10^{-5}$, which should be much nearer the
onset of Regge behavior.
\begin{figure}[hbt]
\begin{center}
\epsfig{file=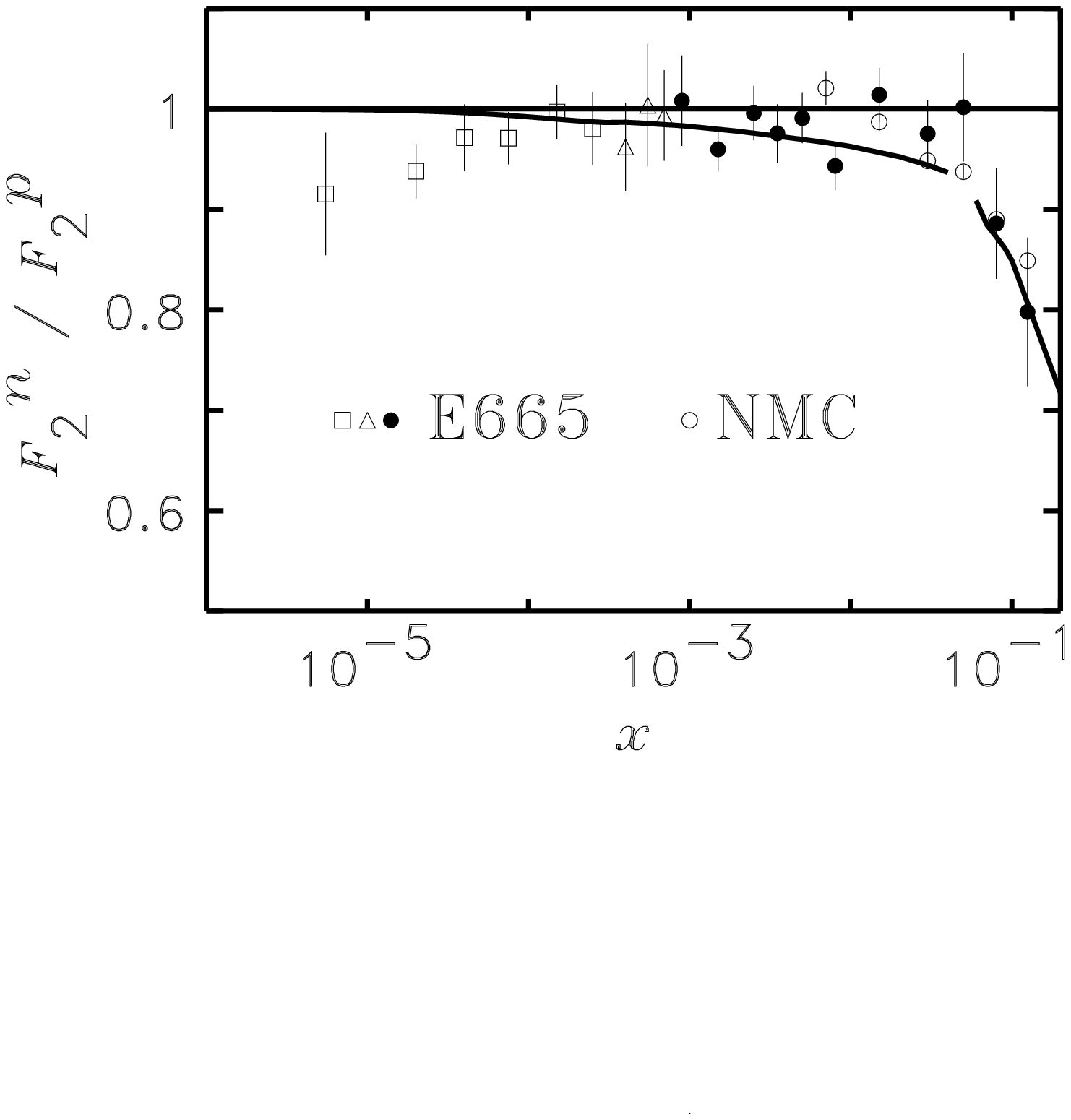,height=7cm}
\caption{$x$ dependence of the $F_2^n/F_2^p$ structure function ratio,
	compared with the values extracted from the E665 \protect\cite{E_D}
	and NMC data \protect\cite{N_D}, and corrected for nuclear shadowing.}
\end{center}
\end{figure}

In Fig.6 we show the ratio $F_2^n/F_2^p$ extracted from the low $x$
E665 Collaboration data \cite{E_D}, as well as the earlier NMC data at
larger $x$ \cite{N_D}, which have been corrected for the effects of
shadowing and MEC \cite{MTQ} (note that the E665 data are not taken
at fixed $Q^2$).
With these corrections, the data are in fact consistent with the hypothesis
that $F_2^n = F_2^p$ in the small $x$ region (say $x < 10^{-3}$).

\subsection{The Gottfried Sum Rule}  

The main interest in the NMC measurement of the neutron structure function 
at low $x$ was an accurate test of the Gottfried sum rule:
\begin{eqnarray}
S_G
&=& \int_0^1 dx\ \frac{ F_2^p(x) - F_2^n(x) }{ x } \\
&=& {1 \over 3} - \int_0^1 dx\ (\bar d(x) - \bar u(x)).
\end{eqnarray}
In the naive quark model, where $\bar d = \bar u$, this gives $S_G = 1/3$. 
Ignoring nuclear effects, the most recent experimental value obtained by 
the NMC (over the range 0.004 to 0.8) is $S_G^{\rm exp} 
= 0.2281 \pm 0.0065$ \cite{KAB}, indicating a clear violation of SU(2) 
flavor symmetry in the proton's sea.
However, because the structure function difference in $S_G$ is weighted 
by a factor $1/x$, any small differences between $F_2^n$ and
$F_2^{n \rm (bound)}$ are amplified for $x \rightarrow 0$.

Including the nuclear corrections, the overall effect on the experimental
value for $S_G$:
\begin{eqnarray}
S_G
&=& S_G^{\rm exp}
 + \int_0^1 dx { \delta^{\rm (shad)} F_2^D(x) \over x },   
\end{eqnarray}
is a reduction of between --0.010 and --0.026, or about 4 and 10\% of 
the measured value without the shadowing correction \cite{MTD}.
Therefore a value that reflects the ``true'' Gottfried sum should be
around $S_G \approx 0.2$, which represents some 30\% reduction from
the naive quark model prediction, $S_G = 1/3$.

\subsection{SU(2)$_F$ Violation and the Chiral Structure of the Nucleon}

The fundamental degrees of freedom in QCD are quarks and gluons, and for
a considerable time there was great reluctance to include any other
degrees of freedom in modeling hadron structure.
On the other hand, extensive studies of non-perturbative QCD have shown 
that the chiral symmetry of the QCD Lagrangian is dynamically broken and 
that the resulting, massive constituent quarks must be coupled to pions.
As pseudo-Goldstone bosons, the latter would be massless in the chiral 
limit.
Most importantly, as emphasised by extensive work on chiral perturbation 
theory, {\em no perturbative treatment of $q \bar{q}$ creation and 
annihilation can ever generate the non-analytic behaviour in the light
quark mass for physical quantities (such as $M, <r^2>, \sigma_{\pi N}$) 
which are generated by these Goldstone bosons}.
As a consequence, it is now difficult to imagine a realistic quark model
which does not incorporate at least the pion cloud of the nucleon.
\begin{figure}[htb]
\begin{center}
\epsfig{file=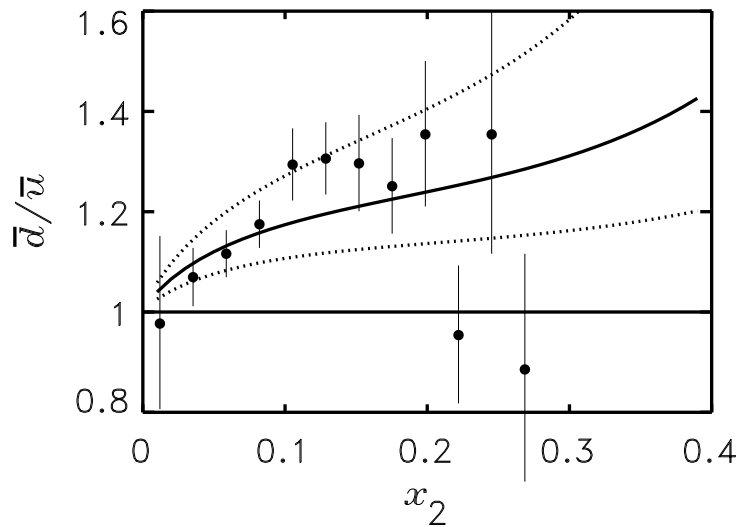,height=8cm}
\caption{Comparison of the value of $\bar{d}/\bar{u}$ extracted from the
	recent E866 data \protect\cite{E866} with the expectations from
	just the $\pi N$ Fock component of the nucleon wave function,
	for three different form factors at the $\pi N$ vertex.}
\end{center}
\end{figure}

The early development of chiral quark models began in the late 1970's
\cite{CBM,IM,CHT,BRR} and it became clear that the pion field had very
important practical consequences for the low energy properties of hadrons.
For example, the coupling to a ``bare'' nucleon lowers its mass by several
hundred MeV, and can contribute a large part of the $N-\Delta$ mass 
difference, while the charge form factor of the neutron is a first
order effect of its pion cloud \cite{CBM}.

However, the relevance of the pion cloud for deep inelastic scattering,
which had been first realized by Feynman and Sullivan \cite{SULL},
was not explored in depth until the discovery of the nuclear EMC effect,
when it was suggested that the effect might be caused by a nuclear
enhancement of the pion field of the bound nucleons \cite{CLLS,ET}.
At this time it was realized that the light mass of the pion would lead
to an enhancement of the non-strange over the strange sea of the nucleon
\cite{THO83}, which was used to put a constraint on the pion--nucleon
form factor --- a limit that has since been explored in detail \cite{FMS}.
A more important consequence of the nucleon's pion cloud, which was also 
pointed out in Ref.\cite{THO83}, was an excess of $\bar{d}$ over $\bar{u}$
quarks.
In particular, simple Clebsch-Gordan coefficients for isospin show that
the pion cloud of the proton is in the ratio 2:1 for $\pi^+:\pi^0$.
Since the $\pi^+$ contains only a valence $\bar{d}$ and the $\pi^0$
equal amounts of $\bar{d}$ and $\bar{u}$, this component of the pion
cloud of the nucleon yields a ratio for $\bar{d}:\bar{u}$ of 5:1.

On the other hand, perturbative QCD inevitably leads to the conclusion
that $\bar{u} = \bar{d}$, a result known as SU(2)-flavor, SU(2)$_F$, 
symmetry.
The latter is a very misleading name as there is {\em no rigorous
symmetry involved}.
Indeed, the prediction that the component of the sea arising from
the long-range piece of the pion cloud of the nucleon satisfies 
$\bar{d}:\bar{u}$ = 5:1 is totally consistent with charge independence.
Thus the violation of the Gottfried sum rule and the subsequent
measurement of $\bar{d} - \bar{u}$ gives us direct evidence that there
is a sizable non-perturbative component of the nucleon sea.
Various studies of the pion cloud of the nucleon since the original NMC
measurement \cite{NMCO} have concluded that this is indeed the most
likely explanation of the observed violation of the sum rule 
\cite{SPTH,HM,SST,KL,HSB}.

In Fig.7 we show the calculated ratio of $\bar{d}/\bar{u}$ from the
$\pi N$ Fock component of the nucleon wave function as a source of 
asymmetry, in comparison with the data points extracted from the 
preliminary results \cite{E866} from the E866 collaboration on the
ratio of $pD$ and $pp$ Drell-Yan cross sections.
The calculation was performed in the light-cone formalism \cite{MTLC,ZOLL}
with a monopole $\pi NN$ form factor mass parameter $\Lambda = 0.7,
1.0$ and 1.3 GeV --- from smallest to largest.
Clearly the agreement is qualitatively excellent, but a more detailed
analysis needs to be carried out once all the data have been analysed.
Let us emphasise once more the importance of these data, which are giving
us direct insight into the way dynamical chiral symmetry breaking is
realized in the nucleon.

\subsection{Other Non-perturbative Components of the Sea}

The experience with the breaking of SU(2)$_F$ symmetry which we have
just described leads us to take much more seriously the possibility of
an intrinsic component of the strange and charm quark sea.
The former was first discussed by Signal and Thomas \cite{ST} and has
recently been investigated by a number of authors \cite{BROD,SMSBAR}
in the light of new neutrino data from CCFR \cite{CCFR}.
This will clearly be the subject of much more detailed investigation in 
future.
With regard to the question of intrinsic charm \cite{BROD,MT_HERA},
we refer to our earlier discussion of the HERA anomaly for some sense
of the issues involved.
This too will clearly be a subject of intense investigation over the next
few years.

\section{CONCLUSION}

At this stage it should be clear that the study of deep-inelastic
scattering from few-body systems, notably the deuteron, provides a rich
source of new information on hadron structure. It is a fascinating
source of challenges for few-body physicists, including such issues as
relativity, off-shell effects, meson exchange currents and vector meson
dominance.

As the result of a careful analysis of binding effects in deep-inelastic 
scattering from the deuteron, there has been a dramatic change in the 
experimental value of the $d/u$ ratio in the valence region.
It seems to be consistent with the perturbative QCD prediction that
$d/u \ra 1/5$ as $x \ra 1$, rather than 0 --- as assumed in all standard
parton distributions (and text books).
This analysis strongly indicates a need for alternative determinations of 
$d/u$ --- e.g. through semi-inclusive measurements --- just to be sure.
Meanwhile, there is an urgent need to work out the quantitative
consequences for this finding in the kinematic region relevant
to the HERA anomaly.

It was reassuring that the model dependent analysis of the  
``non-convolution'' corrections in the deuteron gave values of order 1--2\%.
However, it remains an open question how big they could be in heavier nuclei.
In particular, it would be of great interest to have a relativistic 
$^3$He-N vertex of the Gross type.
This is even more important because $^3$He is usually used in the 
determination of $g_1^n$.

In the small-$x$ region a careful analysis of recent CERN and Fermilab
data in terms of VMD, Pomeron and meson exchange corrections suggests 
that $F_2^n$ and $F_2^p$ are equal below $x \sim 10^{-3}$.
A very important discovery in this region has been the violation of the
Gottfried sum rule.
The physics involved in this violation has been made explicit by recent
data from the E866 Collaboration at Fermilab, which shows the $x$-dependence
of the difference between $\bar{d}$ and $\bar{u}$.

The theoretical origin of the $\bar{d}$--$\bar{u}$ difference seems to 
be the pion cloud of the nucleon, which in turn is a direct consequence 
of dynamical chiral symmetry breaking.
Further study of the shape of this component of the intrinsic sea of the 
nucleon should throw new light on the way chiral symmetry is realized in 
non-perturbative QCD. From another 
point of view, the verification of this prediction for the 
light quark sea leads us to be bolder in looking for differences between 
$s$ and $\bar{s}$, and $c$ and $\bar{c}$.

To conclude, we have seen that deep inelastic scattering is potentially
a gold mine of information concerning the structure of hadronic matter.
Thus far we have really only begun to scratch the surface.

\begin{center}
{\bf Acknowledgements}
\end{center}

We would like to acknowledge important contributions to the work
described here by A. W. Schreiber and G. Piller. This work was supported
by the Australian Research Council, and the DOE grant DE-FG02-93ER-40762.


\begin{thebibliography}{9}
%
\bibitem{EMC}
J.J. Aubert {\em et al.} (EMC),
Phys. Lett. B {\bf 123} (1983) 275;
Nucl. Phys. {\bf B293} (1987)~740.
%
\bibitem{ARN}
M. Arneodo {\em et al.},
Phys. Rep. {\bf 240} (1994) 301.
%
\bibitem{GST}
D.F. Geesaman, K. Saito and A.W. Thomas,
Annu. Rev. Nucl. Part. Sci. {\bf 45} (1995)~337.
%
\bibitem{GSRT}
P.A.M. Guichon {\em et al.},
Nucl. Phys. {\bf A601} (1996) 349.
%
\bibitem{SAI}
K. Saito and A.W. Thomas,
Nucl. Phys. {\bf A574} (1994) 659.
%
\bibitem{NMCO}
P. Amaudruz {\it et al.} (NMC),
Phys. Lett. B {\bf 295} (1992) 159;
Phys. Rev. Lett. {\bf 66} (1991)~2712.
%
\bibitem{E866}
E.A. Hawker {\em et al.} (E866 Collaboration),
``Measuring the $\bar u-\bar d$ Asymmetry in the Proton Sea'',
presented at XXXII Moriond Conference, 22-29 March 1997.
%
\bibitem{THO83}
A.W. Thomas,
Phys. Lett. {\bf 126} B (1983) 97.
%
\bibitem{SPTH}
J. Speth and A.W. Thomas,
``Mesonic Contributions to the Spin and Flavor Structure of the Nucleon''
(J\"ul-3283, Sept. 1996), to appear in Adv. Nucl. Phys. (1997).
%
\bibitem{KM}
D. Kusno and M. Moravcsik,
Phys. Rev. D {\bf 20} (1979) 2734.
%
\bibitem{AW}
W.B. Atwood and G.B. West,
Phys. Rev. D {\bf 7} (1973) 773.
%
\bibitem{BR}
A. Bodek and J.L. Ritchie,
Phys. Rev. D {\bf 23} (1981) 1070.
%
\bibitem{FS}
L.L. Frankfurt and M.I. Strikman,
Phys. Lett. B {\bf 76}  (1978) 333;
Phys. Rep. {\bf 76} (1981) 215.
%
\bibitem{KU}
L.P. Kaptari and A.Yu. Umnikov,
Phys. Lett. B {\bf 259} (1991) 155.
%
\bibitem{LG}
F. Gross and S. Liuti,
Phys. Rev. C {\bf 45} (1992) 1374;
%
\bibitem{BT}
M.A. Braun and M.V. Tokarev,
Phys. Lett. B {\bf 320} (1994) 381.
%
\bibitem{DT}
G.V. Dunne and A.W. Thomas,
Nucl. Phys. {\bf A455} (1986) 701.
%
\bibitem{NW}
K. Nakano and S.S.M. Wong,
Nucl. Phys. {\bf A530} (1991) 555.
%
\bibitem{FS88}
L.L. Frankfurt and M.I. Strikman,
Phys. Rep. {\bf 160} (1988) 235.
%
\bibitem{MILL}
H. Jung and G.A. Miller,
Phys. Lett. B {\bf 200} (1988) 351.
%
\bibitem{CONV}
R.L. Jaffe,
in {\em Relativistic Dynamics and
Quark-Nuclear Physics},
eds. M.B.Johnson and A.Pickleseimer
(Wiley, New York, 1985).
S.V. Akulinichev, S.A. Kulagin and G.M. Vagradov,
Phys. Lett. B {\bf 158} (1985) 485;
S.A. Kulagin, G. Piller and W. Weise,
Phys. Rev. C {\bf 50} (1994) 1154.
%
\bibitem{MST1}
W. Melnitchouk, A.W. Schreiber and A.W. Thomas,
Phys. Rev. D {\bf 49} (1994) 1183.
%
\bibitem{BC}
R. Blankenbecler and L.F. Cook,
Phys. Rev. {\bf 119} (1960) 1745.
%
\bibitem{BG}
W.W. Buck and F. Gross,
Phys. Rev. D {\bf 20} (1979) 2361;
R.G. Arnold, C.E. Carlson and F. Gross,
Phys. Rev. C {\bf 21} (1980) 1426;
F. Gross, J.W. Van Orden and K. Holinde,
Phys. Rev. C {\bf 45} (1992) 2094.
%
\bibitem{REL}
B.D. Kiester and J.A. Tjon,
Phys. Rev. C {\bf 26} (1982) 578;
J.A. Tjon,
Nucl. Phys. {\bf A463} (1987) 157C;
D. Plumper and M.F. Gari,
Z. Phys. A {\bf 343} (1992) 343.
%
\bibitem{MSM}
P. Mulders, A.W. Schreiber and H. Meyer,
Nucl. Phys. {\bf A549} (1992) 498.
%
\bibitem{MST2}
W. Melnitchouk, A.W. Schreiber and A.W. Thomas,
Phys. Lett. B {\bf 335} (1994) 11.
%
\bibitem{WHIT}
L.W. Whitlow {\em et al.},
Phys. Lett. B {\bf 282} (1992) 475.
%
\bibitem{US}
T. Uchiyama and K. Saito,
Phys. Rev. C {\bf 38} (1988) 2245.  
%
\bibitem{LIGHTFRONT}
J. Carbonell, B. Desplanques, V.A. Karmanov and J.-F.Mathiot,
to appear in Phys.~Rep.
%
\bibitem{MTNP}
W. Melnitchouk and A.W. Thomas,
Phys. Lett. B {\bf 377} (1996) 11.
%
\bibitem{BODEK}
A. Bodek {\em et al.},
Phys. Rev. D {\bf 20} (1979) 1471;
A. Bodek and J.L. Ritchie,
Phys. Rev. D {\bf 23} (1981) 1070.
%
\bibitem{F2PAR}
M. Arneodo {\em et al.} (NMC), 
Phys. Lett. B {\bf 364} (1995) 107. 
%
\bibitem{CLO79}
F.E. Close,
{\em An Introduction to Quarks and Partons}
(Academic Press, 1979).
%
\bibitem{CLO73}
F.E. Close,
Phys. Lett. B {\bf 43} (1973) 422.
%
\bibitem{CAR75}
R. Carlitz, 
Phys. Lett. B {\bf 58} (1975) 345.
%
\bibitem{CT}
F.E. Close and A.W. Thomas,
Phys. Lett. B {\bf 212} (1988) 227.
%
\bibitem{FJ}
G.R. Farrar and D.R. Jackson,
Phys. Rev. Lett. {\bf 35} (1975) 1416.
%
\bibitem{BBS}
S.J. Brodsky, M. Burkardt and I. Schmidt,
Nucl. Phys. {\bf B441} (1995) 197.
%
\bibitem{GOMEZ}
J. Gomez {\em et al.},
Phys. Rev. D {\bf 49} (1994) 4348.
%
\bibitem{CDHS}
H. Abramowicz {\em et al.} (CDHS Collaboration),
Z. Phys. C {\bf 25} (1983) 29.
%
\bibitem{H1}
C. Adloff {\em et al.} (H1 Collaboration),
hep-ex/9702012.
%
\bibitem{ZEUS}
J. Breitwig {\em et al.} (ZEUS Collaboration),
hep-ex/9702015.
%
\bibitem{BROD}
S.J. Brodsky and B.-Q. Ma,
Phys. Lett. B {\bf 381} (1996) 317.
%
\bibitem{KLT}
S. Kuhlmann, H.L. Lai and W.-K. Tung,
hep-ph/9704338.
%
\bibitem{MT_HERA}
W. Melnitchouk and A.W. Thomas,
hep-ph/9707387.
%
\bibitem{CHENG}
H.-Y. Cheng,
Int. J. Mod. Phys. A {\bf 11} (1996) 5109;
M. Anselmino {\em et al.},
Phys. Rep. {\bf 261} (1995) 1.
%
\bibitem{EMC_SPIN}
J. Ashman {\em et al.} (EMC),
Phys. Lett. B {\bf 206} (1988) 364.
%
\bibitem{VER}
S.A. Larin, T. van Ritbergen and J.A.M. Vermaseren,
hep-ph/9702435.
%
\bibitem{DPOL}
D. Adams {\em et al.} (SMC),
Phys. Lett. B {\bf 357} (1995) 248;
K. Abe {\em et al.} (E143 Collaboration),
Phys. Rev. Lett. {\bf 75} (1995) 25.
%
\bibitem{MPT}
W. Melnitchouk, G. Piller and A.W. Thomas,
Phys. Lett. B {\bf 346} (1995) 165;
G. Piller, W. Melnitchouk and A.W. Thomas,
Phys. Rev. C {\bf 54} (1996) 894.
%
\bibitem{KMPW}
S.A. Kulagin, W. Melnitchouk, G. Piller and W. Weise,
Phys. Rev. C {\bf 52} (1995) 932.
%
\bibitem{KWBD}
J. Kwiecinski and B. Badelek,
Phys. Lett. B {\bf 208} (1988) 508.
%
\bibitem{KW}
J. Kwiecinski,
Z. Phys. C {\bf 45} (1990) 461.
%
\bibitem{BK}
B. Badelek and J. Kwiecinski,
Nucl. Phys. {\bf B370} (1992) 278;
Phys. Rev. D {\bf 50} (1994)~4.
%
\bibitem{MTD}
W. Melnitchouk and A.W. Thomas,
Phys. Rev. D {\bf 47} (1993) 3783.
%
\bibitem{MTA}
W. Melnitchouk and A.W. Thomas,
Phys. Lett. B {\bf 317} (1993) 437.
%
\bibitem{OTHERD}
V.R. Zoller,
Z. Phys. C {\bf 54} (1992) 425;
%
G. Piller, W. Ratzka and W. Weise,
Z. Phys. A {\bf 352} (1995) 427;
%
H. Khan and P. Hoodbhoy,
Phys. Lett. B {\bf 298} (1993) 181.
%
\bibitem{NMC}
P. Amandruz {\em et al.} (NMC),
Phys. Rev. Lett. {\bf 66} (1991) 2712.
%
\bibitem{CBM}
S. Th\'eberge, G.A. Miller and A.W. Thomas,
Phys. Rev. D {\bf 22} (1980) 2838;
{\em ibid} D {\bf 23} (1981) 2106(e);
%
A.W. Thomas,
Adv. Nucl. Phys. {\bf 13} (1984) 1;
%
G.A. Miller,
Int. Rev. Nucl. Phys. {\bf 2} (1984) 190.
%
\bibitem{HERA}
T. Ahmed {\em et al.},
Phys. Lett. B {\bf 348} (1995) 681.
%
\bibitem{DOLA}  
A. Donnachie and P.V. Landshoff,
Phys. Lett. B {\bf 191} (1987) 309.
%
\bibitem{DLQ}   
A. Donnachie and P.V. Landshoff,
Z. Phys. C {\bf 61} (1994) 139.
%
\bibitem{CLLS}
C.H. Llewellyn Smith,
Phys. Lett. B {\bf 128} (1983) 107.
%
\bibitem{ET}
M. Ericson and A.W. Thomas,
Phys. Lett. B {\bf 128} (1983) 112.
%
\bibitem{KAP} 
L.P. Kaptari, A.I. Titov, E.L. Bratkovskaya and A.Yu. Umnikov,
Nucl. Phys. {\bf A512} (1990) 684.
%
\bibitem{N_D}
M. Arneodo {\em et al.} (NMC),
Phys. Rev. D {\bf 50} (1994) 1.
%
\bibitem{E_D}
M.R. Adams {\em et al.} (E665 Collaboration),
Phys. Rev. Lett. {\bf 75} (1995) 1466;
Phys. Lett. B {\bf 309} (1993) 477.
%
\bibitem{MTQ}
W. Melnitchouk and A.W. Thomas,
Phys. Rev. C {\bf 52} (1995) 3373.
%
\bibitem{KAB}
E. Kabuss (NMC),
hep-ph/9706435.
%
\bibitem{IM}
T. Inoue and T. Maskawa,
Prog. Theor. Phys. {\bf 54} (1975) 1833.
%
\bibitem{CHT}
A. Chodos and C.B. Thorn,
Phys. Rev. D {\bf 12} (1975) 359.
%
\bibitem{BRR}
G.E. Brown and M. Rho,
Phys. Lett. B {\bf 82} (1979) 177.
%
\bibitem{SULL}
J.D. Sullivan,
Phys. Rev. D {\bf 5} (1972) 1732.
%
\bibitem{FMS}
L.L. Frankfurt, L. Mankiewicz and M.I. Strikman,
Zeit. Phys. A {\bf 334} (1989) 334;
%
W. Koepf, L.L. Frankfurt and M.I. Strikman,
Phys. Rev. D {\bf 53} (1996) 2586.
%
\bibitem{HM}
E.M. Henley and G.A. Miller,
Phys. Lett. B {\bf 251} (1990) 453.
%
\bibitem{SST}
A.I. Signal, A.W. Schreiber and A.W. Thomas,
Mod. Phys. Lett. A {\bf 6} (1991) 271.
%
\bibitem{KL}
S. Kumano and J.T. Londergan,
Phys. Rev. D {\bf 44} (1991) 717;
S. Kumano, hep-ph/9702367.
%
\bibitem{HSB}
W.-Y.P. Hwang, J. Speth and G.E. Brown,
Zeit. Phys. A {\bf 339} (1991) 383.
%
\bibitem{MTLC}
W. Melnitchouk and A.W. Thomas,
Phys. Rev. D {\bf 47} (1993) 3794.
%
\bibitem{ZOLL}
V.R. Zoller,
Z. Phys. C {\bf 60} (1993) 141.
%
\bibitem{ST}
A.I. Signal and A.W. Thomas,
Phys. Lett. B {\bf 191} (1987) 205.
%
\bibitem{SMSBAR}
X. Ji and J. Tang,
Phys. Lett. B {\bf 362} (1995) 182;
%
H. Holtmann {\em et al.},
Nucl. Phys. {\bf A569} (1996) 631;
%
W. Melnitchouk and M. Malheiro,
Phys. Rev. C {\bf 55} (1997) 431. 
%
\bibitem{CCFR}
A. Bazarko {\em et al.} (CCFR Collaboration),
Zeit. Phys. C {\bf 65} (1995) 189.
%
\end{thebibliography}
\end{document}